\documentclass[final]{siamltex}

\title{Algorithmic Solutions for Several Offline Constrained Resource Processing and Data Transfer Multicriteria Optimization Problems}

\author{Mugurel Ionu\c t Andreica\thanks{Computer Science and Engineering Department, Faculty of Automatic Control and Computer Science, Bucharest, Romania ({\tt mugurel.andreica@cs.pub.ro}).} \and Nicolae \c T\u apu\c s \thanks{Computer Science and Engineering Department, Faculty of Automatic Control and Computer Science, Bucharest, Romania ({\tt nicolae.tapus@cs.pub.ro}).}}

\begin{document}

\maketitle

\begin{abstract}
In this paper we present novel algorithmic solutions for several resource processing and data transfer multicriteria optimization problems. The results of most of the presented techniques are strategies which solve the considered problems (almost) optimally. Thus, the developed algorithms construct intelligent strategies which can be implemented by agents in specific situations. All the described solutions make use of the properties of the considered problems and, thus, they are not applicable to a very general class of problems. However, by considering the specific details of each problem, we were able to obtain very efficient results.
\end{abstract}



\pagestyle{myheadings}
\thispagestyle{plain}
\markboth{M. I. ANDREICA AND N. \c T\u APU\c S}{ALGO. SOL.-OFFLINE CONSTR. RES. PROC. \& DATA TRANS. M.-CRIT. OPT. PROB.}

\section{Introduction}

Resource processing and data transfer multicriteria optimization problems occur in many situations, particularly in large distributed systems, like Grids, distributed databases, live and on-demand video streaming applications, peer-to-peer systems, and so on. The problems that occur in practical settings are of a wide diversity and consider the optimization of various parameters, while imposing constraints on others. Handling such problems from a general perspective is useful when trying to understand and classify them, but it is of little use when we want to solve a specific problem. In this paper we consider several such multicriteria optimization problems, for which we provide novel, very specific algorithmic solutions. Instead of presenting a general technique and analyzing its efficiency on various types of optimization problems, we focus on the problems and develop different solutions for each problem. The considered problems are mainly offline, meaning that all the required information is available in advance. From this point of view, the presented results cannot be applied directly in real-time settings (where most of the resource processing and data transfer optimization problems occur). However, from a theoretical point of view, the developed algorithms are of significant interest and they are the first steps towards obtaining efficient online solutions to some of the considered problems.

The rest of this paper is structured as follows. In Section 2 we discuss several offline (multi)point-to-(multi)point data transfer optimization problems, like finding a deadline-constrained packet transfer strategy, computing a minimum cost communication spanning tree when the link providers have special offers, and finding a subset of edges (vertices) of maximum average weight (when we have two weights assigned to every edge or vertex). In Section 3 we propose a new agent-based peer-to-peer content delivery architecture. We present generic methods and guidelines and discuss the issues of point-to-point and multicast data transfers within the topology. In Section 4 we discuss several constrained (multi-)permutation construction and sorting problems. Such problems are partially related to efficiently ordering the packets of a communication flow in the transmission buffer. In Section 5 we present new algorithmic extensions to the Union-Find and Split-Find problems. These are set maintenance problems which require efficient management and querying of sets of elements. They have applications in several resource processing situations (e.g. when reserving resources, see \cite{centr}). In Section 6 we discuss another resource processing problem, for which we present an optimal algorithm. The problem consists of activating/deactivating resources in a graph with bounded treewidth by using a minimum cost strategy. This problem is a more general version of the \emph{minimum all-ones problem} \cite{min_all_ones}, which assumes that the final state of each resource must always be \emph{active} and the cost of selecting a resource is always $1$. Finally, in Section 7 we discuss related work and in Section 8 we conclude.

\section{Offline (Multi)Point-to-(Multi)Point Data Transfer Optimization}

\subsection{Optimal Deadline-Constrained Packet Transfer Strategy}

We consider a directed graph with $n$ vertices and $m$ edges. A packet is sent from the source node $s$ at time $0$ and must reach the destination node $d$ by time $T$. Every directed edge $e$ from a vertex $u$ to a vertex $v$ has an associated start time $tstart(e)$ and a finish time $tfinish(e)$ ($tfinish(e)>tstart(e)$). The meaning of these parameters is that the packet can only be sent along that edge, from $u$ to $v$, starting at the moment $tstart(e)$ and will only arrive at vertex $v$ at the moment $tfinish(e)$. Thus, edge $e$ corresponds to a reservation in the underlying network. Moreover, out of the total packet transmission time (equal to $tfinish(e)-tstart(e)$), $twait(e)$ is the total time during which the packet has to wait in the waiting queues (e.g. it must wait for some data processing task or must wait until other packets before it are sent along the edge). The time between the moment when the packet arrives to a vertex $u$ and the moment when it is sent to another vertex $v$ (or between the moment when it last arrives at vertex $d$ and the moment $T$) also counts as waiting time. We want to find a packet transfer strategy minimizing the total waiting time. The steps of the algorithm are described below.

\begin{enumerate}
\item For every vertex $v$ of the graph:\begin{enumerate}
\item We will sort together the incoming and outgoing edges in increasing order, according to a weight assigned to every edge: for an incoming edge $e$ from a vertex $u$ to vertex $v$, the weight is $w(v,e)=tfinish(e)$; for an outgoing edge $e$ from vertex $v$ to a vertex $u$, the weight is $w(v,e)=tstart(e)$.
\item If two edges (an incoming and an outgoing one) have the same weight, then we will place the incoming edge before the outgoing edge in the sorted order.
\item For every edge in the sorted order of a vertex $v$ we will store its type: incoming or outgoing.
\item Let $deg(v)$ be the total number of incoming and outgoing edges adjacent to vertex $v$.
\end{enumerate}
\item We will compute $TWmin(v,i)$=the minimum total waiting time required for the packet to be located at vertex $v$ at the time moment $tm(v,i)$=the weight of the $i^{th}$ edge in the sorted order for vertex $v$ ($1\leq i\leq deg(v)$).
\item We will consider $TWmin(v,0)=+\infty$ and $tm(v,0)=0$ for every vertex $v\neq s$ and $tm(s,0)=TWmin(s,0)=0$.
\item We will sort ascendingly all the time moments $tm(v,i)$ ($i\geq 1$) in increasing order (e.g. by merging the lists of time moments $tm(*,*)$) and we will store for each moment the associated values $v$ and $i$.
\item We will traverse all the time moments $tm(v,i)$ in increasing order.\begin{enumerate}
\item If $tm(v,i)$ corresponds to an incoming edge $e$ (from a vertex $u$ to vertex $v$), then we will first find the index $j$ of the edge $e$ in the sorted order of the edges adjacent to vertex $u$. 
\item We will have $TWmin(v,i)=min\{TWmin(v,i-1)+tm(v,i)-tm(v,i-1)$, $TWmin(u,j)+twait(e)\}$.
\item If $tm(v,i)$ corresponds to an outgoing edge $e$ from vertex $v$ to a vertex $u$ then we set $TWmin(v,i)=TWmin(v,i-1)+tm(v,i)-tm(v,i-1)$.
\end{enumerate}
\end{enumerate}

We can find the index $j$ of an edge $e$ in the sorted order of a vertex $u$ by using a hash table $HT(u)$. After sorting the edges adjacent to vertex $u$ we traverse these edges: let the edge $e'$ be the $i^{th}$ edge in this order - then we insert the pair $(key=e', value=i)$ in $HT(u)$. Thus, in order to find the index $j$ associated to an edge $e$ in the sorted order of a vertex $u$ we just search in $HT(u)$ the value associated to the key $e$. Such a lookup takes $O(1)$ time and the overall time complexity is $O(m\cdot log(m))$.

The final answer is $min\{TWmin(d,i)+T-tm(d,i)|1\leq i\leq deg(d),tm(d,i)\leq T\}$.

As we can notice, the problem can also be interpreted as a shortest path problem in the graph of the pairs $(v,i)$, where the starting pair is $(s,0)$. We have an edge from each pair $(v,i-1)$ to the pair $(v,i)$ with cost $tm(v,i)-tm(v,i-1)$ ($1\leq i\leq deg(v)$). Moreover, we have an edge from each pair $(u,j)$ to a pair $(v,i)$, with cost $twait(e)$ if $e$ is an edge from $u$ to $v$ and is the $j^{th}$ edge in the sorted order of vertex $u$ and the $i^{th}$ edge in the sorted order of vertex $v$. With this interpretation, we can compute a shortest path from $(s,0)$ to the pairs $(d,*)$, in $O(m\cdot log(m))$ time (the graph of pairs has $O(m)$ vertices and edges). If we denote by $TWmin(v,i)$=the length of the shortest path from $(s,0)$ to $(v,i)$, then the answer is computed the same as before.

\subsection{Maximum (Minimum) Ratio Constrained Subsets of Edges (Vertices)}

We are given a (directed) graph $G$ with $n$ vertices and $m$ edges. Each (directed) edge $(u,v)$ has two associated values: $p(u,v)\geq 0$ and $q(u,v)>0$, and each vertex $u$ has two associated values: $p(u)\geq 0$ and $q(u)>0$ (e.g. these values could be bandwidth, cost, or latency). We want to find a subset of edges $E$ (vertices $V$) having a specified property $Prop$, such that its ratio is maximum (minimum). The ratio $A$ of a subset of edges $E$ (vertices $V$) is defined as $\frac{\sum_{(u,v)\in E}{p(u,v)}}{\sum_{(u,v)\in E}{q(u,v)}}$ ($\frac{\sum_{u\in V}{p(u)}}{\sum_{u\in V}{q(u)}}$). We will use a technique which was previously discussed in other research papers \cite{eppstein_avg} for solving other optimization problems. Let's consider the following problem: We are given a (directed) graph $G$ with $n$ vertices and $m$ edges. Each (directed) edge $(u,v)$ (vertex $u$) has a weight $w(u,v)$ ($w(u)$), which may be zero, positive or negative. We want to find a subset of edges $E$ (vertices $V$) satisfying property $Prop$, such that the sum of the weights of the edges $(u,v)\in E$ (vertices $u\in V$) is maximum (minimum). We will denote the optimization algorithm by $OPTA(G)$, where $G$ is the graph. $OPTA(G)$ returns the weight of the subset of edges (vertices). The algorithm will compute several other values, based on which we will easily be able to find the subset of edges (vertices) itself (not just its weight). With this algorithm, we will be able to solve the maximum (minimum) ratio problem as follows. We will binary search the maximum (minimum) ratio $A_{opt}$ in the interval $[0,AMAX]$, where $AMAX$ is a good upper bound (e.g. $AMAX$ is the maximum among the $p(*,*)$ ($p(*)$) values, divided by the minimum among the $q(*,*)$ ($q(*)$) values). For a candidate value $A_{cand}$, we will construct a graph $G'$ having the same set of edges and vertices. The weight of a (directed) edge $(u,v)$ (vertex $u$) in $G'$, $w(u,v)$ ($w(u)$), will be $p(u,v)-A_{cand}\cdot q(u,v)$ ($p(u)-A_{cand}\cdot q(u)$). We will now call $OPTA(G')$ and store its value as $A_{res}$. If $A_{res}<0$, then $A_{cand}$ is larger than the maximum (minimum) ratio and we need to consider smaller values; otherwise, $A_{cand}\leq A_{opt}$ and we can test larger values. The binary search ends when the length of the search interval becomes smaller than $\varepsilon$ (where $\varepsilon>0$ is a very small constant). We will now provide a few examples of optimization problems and algorithms ($OPTA$) which can be fit into the generic framework introduced earlier.

Given a directed graph $G$ with $n$ vertices (numbered from $1$ to $n$) and $m$ edges, where each directed edge $(u,v)$ has a weight $w(u,v)$ (which may be zero, negative or positive) and a length $l(u,v)>0$, we want to find a (possibly self-intersecting) path (cycle) of total length at least $L$ and at most $U$, whose total weight is maximum. We will present a pseudo-polynomial algorithm, for the case when the edge lengths are integers. For the path case, we will compute for each vertex $i$ and each number of edges $k$, the value $W_{max}(i,k)$=the maximum total weight of a path with length $k$, which ends at vertex $i$. We have $W_{max}(i,0)=0$ and for $1\leq k\leq U$, we have $W_{max}(i,k)=max\{W_{max}(j,k-l(j,i))+w(j,i)|(j,i)\in G,l(j,i)\leq k\}$ or $-\infty$, if no edge $(j,i)$ with $l(j,i)\leq k$ exists. The maximum total weight of an optimal path is $max\{W_{max}(i,k)|1\leq i\leq n, L\leq k\leq U\}$. The time complexity is $O((n+m)\cdot U)$. For the cycle case, we consider every possible starting vertex $s$ and run the algorithm described above, except that $W_{max}(s,0)=0$ and $W_{max}(i\neq s,0)=-\infty$. The weight of the optimal cycle starting at vertex $s$ is $max\{W_{max}(s,k)|L\leq k\leq U\}$. When there is no constraint on the length of the path (cycle), we first need to decide if the graph contains a cycle whose total weight is positive. If it does, then the weight of the optimal path (cycle) is $+\infty$. We will perform this checking by computing a value $W_{max}(i)$ for each vertex $i$, using the Bellman-Ford-Moore algorithm. We initialize $W_{max}(*)=0$ and insert all the vertices into a queue $Q$. Afterwards, we repeatedly extract from $Q$ the first vertex $i$, consider all the directed edges $(i,j)$ (from $i$ to $j$) and update $W_{max}(j)$. If $W_{max}(i)+w(i,j)>W_{max}(j)$, then we set $W_{max}(j)=W_{max}(i)+w(i,j)$ and insert the vertex $j$ at the end of the queue (if it is not already in the queue). We will also maintain a counter $nin(i)$ for each vertex $i$, denoting the number of times vertex $i$ was inserted into the queue. If, at some point, we have $nin(i)>n$ for some vertex $i$, then a cycle with total positive weight exists in the graph. If the algorithm ends without finding a positive cycle, then the maximum total weight of a cycle is $0$ and that of a path is $max\{W_{max}(i)\}$. The time complexity of the algorithm in this case is $O(n\cdot m)$.

When the graph is a directed path, i.e. we have only the edges $(i,i+1)$ ($1\leq i\leq n-1$), the optimal path problem can be solved with a better time complexity. The unconstrained version is equivalent to computing a maximum sum segment in the sequence of $n-1$ weights $w(i,i+1)$. The length constrained version is equivalent to the length constrained maximum sum segment problem.

\subsection{Minimum Cost Spanning Tree with Special Offers}

We consider here a minimum cost spanning tree problem, augmented with special offers. We have an undirected graph with $n$ vertices (numbered from $1$ to $n$) and $m$ edges. Each edge $e$ connects two different vertices $a(e)$ and $b(e)$, has an owner $o(e)$ and two prices: $np(e)$ and $sp(e)$. $np(e)$ is the normal price of the edge and $sp(e)$ is the special price of the edge; $sp(e)\leq np(e)$. There are $q$ owners overall, numbered from $1$ to $q$. Each owner has a special offer: it allows us to pay the special prices for any edges we want owned by that owner, with the condition that we do not take the special offer of any of the other owners. We want to establish a minimum cost spanning tree of the graph, by using at most one special offer of one of the edge owners.

At first, we will compute the minimum spanning tree of the graph considering normal prices for all the edges (in $O((m+n)\cdot log(n))$ or $O(m+n\cdot log(n))$ time). Let $MSTN$ be the set of $n-1$ edges composing the "normal" minimum spanning tree and let $CMSTN$ be the cost of the minimum spanning tree considering the normal prices.

We will now consider every owner $i$ (from $1$ to $q$) and compute the minimum spanning tree in the case when we take advantage of owner $i$'s special offer, i.e. when we consider the special prices for all the edges owned by $i$. We can recompute each such spanning tree in the same time complexity as when we computed the first minimum spanning tree, but the overall time complexity would be $O(q\cdot (m+n)\cdot log(n))$ or $O(q\cdot (m+n\cdot log(n)))$. Instead, we will proceed as follows. When computing the minimum spanning tree for the special offer of the owner $i$, we will consider only the subset of edges $SE(i)$ composed of all the edges owned by $i$ (for which we consider their special prices) and all the edges from $MSTN$ which are not owned by $i$, for which we consider their normal price. Then, we will compute the minimum spanning tree using only the edges in $SE(i)$. Note that any edge outside of $SE(i)$ cannot be part of this minimum spanning tree. Thus, the time complexity for one minimum spanning tree computation is $O((|SE(i)|+n)\cdot log(n))$ or $O(|SE(i)|+n\cdot log(n))$.

The sum of the values $|SE(i)|$ $(1\leq i\leq q)$ is at most $m+(n-1)\cdot q$. Thus, the overall time complexity will be $O((m+n\cdot q)\cdot log(n))$ or $O(m+n\cdot q+n\cdot q\cdot log(n))$, which is much better than that of the trivial algorithm.

For a more in-depth analysis of this problem (which considers several other constraints and other types of minimum spanning tree algorithms), see \cite{tech_optimiz_comm} (a book based on the first author's Ph.D. thesis).

\section{Towards an Agent-Based Peer-to-Peer Content Delivery Framework}

In this section we propose a novel architecture of an agent-based peer-to-peer content delivery framework in which data transfer optimization techniques can be used. The problems posed by this framework are mainly online, but we will also discuss a semi-offline problem in the last subsection. Note that we are only presenting the generic principles and guidelines in this section, and not a full implementation.

Each agent (which will be called {\em peer} from now on) of the peer-to-peer framework has an identifier which is a point in a d-dimensional space (in which the range of the coordinates in each dimension $i$ is $[0,VMAX(i)]$). The identifiers can be self-generated (e.g. by using hash functions which generate unique values with high probability), or may be computed according to some specific rules (e.g. they may be mapped in the latency space) \cite{netw_dist}. The peers interconnect in a peer-to-peer topology by using a distributed decision making mechanism. In \cite{p2p_mpath} we presented a generic framework for constructing a peer-to-peer topology based on the peers' geometric coordinates. Several methods for choosing neighbors are also described in \cite{p2p_mpath}, out of which we would like to highlight the {\em Hyperplanes} method. This method makes use of a set of hyperplanes, all of which contain the origin. These hyperplanes form $Q$ disjoint regions, whose union is the entire hyper-space. When a peer $X$ has to choose its neighbors from a set $I(X)$, it will divide the peers $Y$ from $I(X)$ according to which region (among the $Q$ regions) they belong to (when peer $X$ is considered to be the origin) and then it will select as neighbors the closest $K$ peers from each region ($K$ may be different for each region and for each peer). Closeness may be defined using any geometric norm $L_{h}$ ($1\leq h\leq +\infty$). The peers periodically broadcast their existence a number of hops away in the topology and the set $I(X)$ is composed of those peers $Y$ which broadcasted their existence to $X$ recently. The topology changes as the sets $I(*)$ change and the peers select new neighbors, until it reaches an equilibrium (if no new peer joins the system and no old peer leaves it).

In order for the peer-to-peer topology construction mechanism to work properly, we assumed that all the peers are connectable (i.e. they have public IP addresses and are not located behind a firewall). Note that the system may still work if, whenever a connectable peer $X$ selects an unconnectable peer $Y$ as its neighbor, $Y$ also selects $X$ as a neighbor. In order to handle unconnectable peers we consider assigning coordinates at the edge of the multidimensional space to these peers. This way we expect that an unconnectable peer $Y$ is less likely to be selected as a neighbor by a connectable peer $X$ which is not selected back as a neighbor by $Y$. This does not solve the problem of two unconnectable peers which may select each other as neighbors. We assume that every peer can detect if it is unconnectable (this may be easier when the peer has a private IP address than when it is located behind a firewall).

In order to send data from one peer to another, we can employ geometric routing (i.e. we repeatedly forward the data to a neighbor which is closer in the geometric space to the destination), or we can use multi-path routing mechanisms \cite{p2p_mpath}.

Searching for a piece of content is not within the scope of this paper. Nevertheless, when the content units have a set of index properties, the approach from \cite{p2p_rquery} can be used in order to obtain references to every content unit (or at most a number $M$ of them) whose properties are within a specified multidimensional range. Once we have a reference to a content unit, we can initiate a data transfer from its owner.

Except for point-to-point data transfer services, we are interested in supporting multicast communication services within the peer-to-peer architecture. In the following two subsections we will propose two different methods for achieving this. We will first present a generic method for constructing multicast trees on demand. Then, we will discuss the problem of maintaining a multicast tree with improved stability, when we have extra information regarding the peers' life times.

\subsection{On Demand Multicast Data Distribution}

In order to support multicast communication services, we propose building a multicast tree on demand on top of the peer-to-peer topology. Many spanning tree construction algorithms for arbitrary topologies have been proposed in the literature \cite{mcast1,mcast2}. However, the topology proposed in the previous section has a special geometric structure. Thus, we propose the following high-level construction method.

One of the peers $A$ will initiate the multicast data transfer. Let's denote by $Z(X)$ the zone for which a peer $X$ will be responsible. $Z(A)$ will be the entire geometric space. Let's assume that a peer $C$ received a message with the value of $Z(C)$ (peer $A$ is considered to receive such a virtual message in the beginning). Then, $C$ will select a subset of its neighbors $S(C)$ which are located within $Z(C)$ and will construct tree edges from itself towards the peers $B$ in $S(C)$. Then, a disjoint zone $Z(B)$ of the space is computed by peer $C$ for each peer $B$ in $S(C)$, such that: $(1)$ every zone $Z(B)$ is fully contained in $Z(C)$ ; $(2)$ every peer $D$ whose coordinates are within $Z(C)$ (except for $C$ and the peers from $S(C)$) must be located within one of the zones $Z(B)$ ($B \in S(C)$). When using the Hyperplanes neighbor selection methods, the zones $Z(B)$ can be computed by clipping the zone $Z(C)$ with a subset of hyper-planes (thus, the zones are convex). Then, peer $C$ sends the message with $Z(B)$ to every peer $B$ from $S(C)$. The time complexity of this algorithm is optimal, being proportional to the diameter of the underlying peer-to-peer topology.

In some cases, we may decide that the zones $Z(B)$ of the peers $B \in S(C)$ do not have to be disjoint (but their union may be equal to $Z(C)$). Non-disjointness may imply that some peers will end up receiving the data multiple times. This is acceptable if a peer can detect that the received data is duplicate (i.e. the data packets have unique identifiers) and if the complexity of computing non-disjoint zones $Z(B)$ would be too high (either from a computational point of view, or because the description complexity of $Z(B)$ would be too high).

The multicast tree constructed by this method would stay alive only as long as data is transferred along its edges (each edge would have its own expiration timer). A new tree would be constructed for every multicast flow. Although this generates a tree construction overhead for each flow, it is more robust in the presence of node failures (because the trees are not long-lived, they are less likely to experience node failures or departures).

\subsection{Constructing Multicast Trees with Enhanced Stability based on Availability Information}

In this subsection we consider the following scenario. Let's assume that every peer $X$ that enters the peer-to-peer system knows the time moment $T(X)$ when it will leave the system (and can guarantee that it will stay within the system until the moment $T(X)$). We would like to use this information in order to construct and maintain a multicast tree with improved stability.

A method of achieving this is the following. The coordinates of a peer $X$ will be of the form $(v(1), \ldots, v(d-1), v(d)=T(X))$ (where $v(1\leq i\leq d-1)$ can be generated the way we discussed earlier). Then, the peer $X$ will join the peer-to-peer topology. Once its neighbors are stable (i.e. peer $X$ does not select new neighbors for at least $KR(X)$ consecutive gossiping periods), peer $X$ will choose one of its neighbors $Y$ from the peer-to-peer topology as its neighbor in the tree. We will refer to $Y$ as peer $X$'s preferred neighbor ($P(X)=Y$). The set of candidate peers $PC(X)$ from which $P(X)$ can be chosen consists of those peers $Y$ which are neighbors with $X$ in the topology and have $T(Y)\geq T(X)$. If $X$ has no neighbor $Y$ with $T(Y)\geq T(X)$ then $X$ has the highest value of $T(X)$ from the system. In this case, $PC(X)$ consists of all of $X$'s neighbors within the peer-to-peer topology.

The preferred neighbor $P(X)$ may be chosen arbitrarily from $PC(X)$, but we could also use other rules. For instance, for each peer $Y$, we may impose an upper bound $UB(Y)$ on the number of peers $Z$ for which $P(Z)=Y$. Then, we need to remove from $PC(X)$ those peers $Y$ which have already reached their limit. Among the remaining peers we could choose the one with smallest maximum distance from itself towards some other peer in the tree (the distance is equal to the number of tree edges which need to be traversed from one peer to another), in order to minimize the tree diameter. Such information can be maintained through gossiping in the multicast tree (see, for instance, \cite{mcast3}). If all the peers in $PC(X)$ have reached their upper limit, then we may decide to either break one of these limits (e.g. by choosing the peer $Y\in PC(X)$ with the smallest degree) or we could replace a tree edge $Z-Y$ (where $Y\in PC(X)$). The tree edge would be replaced by the edges $Z-X$ and $X-Y$. In this case, it might be better to choose that tree edge $Z-Y$ for which the maximum distance from $Z$ towards a peer in its subtree ($Dmax(Y,Z)-1$ if we use the notations from \cite{mcast3}) is minimum (also in order to try to minimize the diameter). Replacing these edges also implies replacing the corresponding preferred neighbors: if $P(Z)=Y$ then we will set $P(Z)=X$ and $P(X)=Y$; otherwise, we will set $P(Y)=X$ and $P(X)=Z$.

With this method of constructing the tree, the peers $X$ with smaller values of $T(X)$ will be located towards the edge of the tree. Thus, when a peer leaves the system, it will be a leaf in the tree and its departure will not disconnect the tree. This saves us from the need of taking special measures in case of unexpected peer departures which may leave the multicast tree disconnected.

A situation in which the peers $X$ may know exactly the values $T(X)$ is in the case of virtual machines (VMs) deployed in Clouds. If a Cloud computing resource is leased for a fixed amount of time in order to run a virtual machine, then the virtual machine may know the amount of time left before it is powered off. Note also that since leasing virtual machines involves a certain Service Level Agreement (SLA), we can count on the fact the the VM will be running for the specified amount of time without interruptions. In a way, we are transferring the reliability of the SLA negotiated with the Cloud service provider in order to construct a reliable (and stable) multicast tree on top of a flexible peer-to-peer topology. For other scenarios, the assumption of knowing the exact time when the peer leaves the system leads to a semi-offline problem.

\section{Construction and Sorting of (Multi-)Permutations}

\subsection{Permutations with Average Values in Between}

We are interested in constructing a permutation with $n$ elements, such that, for every two distinct values $p(i)$ and $p(j)$ ($i<j$) of the same parity, the position $k$ on which the element $(p(i)+p(j))/2$ is located must satisfy the condition: $k<min\{i,j\}$ or $k>max\{i,j\}$, i.e. the average value must not be located in between the two values in the permutation.

An $O(n\cdot log(n))$ divide-and-conquer algorithm is not difficult to obtain. We notice that we can have all the $n/2$ even numbers located before all the $n-n/2$ odd numbers in the permutation. Then, we need to solve the same problem for $n/2$ and $n-n/2$ elements. We can obtain a valid $n$-element permutation as follows. We multiply each number in the $(n/2)$-element solution by $2$, thus obtaining all the even numbers in the set $\{1,\ldots,n\}$. Afterwards, we turn every element $q$ in the $(n-n/2)$-element solution into $2\cdot q-1$ and obtain all the odd numbers in the set $\{1,\ldots ,n\}$. By concatenating the two sequences, we obtain the desired permutation. The time complexity is obviously $O(n\cdot log(n))$, because, at each step, we need to solve two approximately equal problems. The base case is $n=1$, where the only existing permutations is the solution.

We can improve the time complexity to $O(n)$ as follows:\begin{itemize}
\item If $n=4\cdot k$, then we will call the algorithm for $2\cdot k$ and $2\cdot k$.
\item If $n=4\cdot k+1$, then we will call the algorithm for $2\cdot k$ and $2\cdot k+1$. At the next level, however, for the $2\cdot k$-element solution, we will need to solve two $k$-element problems, and for the $2\cdot k+1$-element solution, we will need to solve a $k$-element problem and one $(k+1)$-element problem.
\item For $n=4\cdot k+2$, we will need to solve two $2\cdot k+1$-element problems at the next level.
\item If $n=4\cdot k+3$, then we will call the algorithm for $2\cdot k+1$ and $2\cdot k+2$. At the next level, however, for the $2\cdot k+1$-element solution, we will need to solve one $k$-element problem and one $(k+1)$-element problem, and for the $2\cdot k+2$-element solution, we will need to solve two $(k+1)$-element problems.
\end{itemize}

What we notice is that at every level of the recursion, there are at most two distinct (sub)problems to be solved. Thus, there are only $O(log(n))$ distinct subproblems to solve. We can use the memoization technique. After computing a valid $k$-element permutation, we store it and retrieve it immediately whenever we require it again. Thus, at subsequent calls, the time complexity will be $O(k)$, instead of $O(k\cdot log(k))$. Because the sizes of the problems at each level decrease exponentially, the total sum of the sizes of the computed permutations is $O(n)$. Thus, the overall time complexity is $O(n)$. 

A simpler $O(n)$ solution is to consider the largest value $m=2^k$ such that $m\geq n$. Then, we solve the problem for a permutation with $m$ elements (which always has only one (sub)problem to solve at each recursion level) and we simply remove all the elements $q>n$ from the obtained permutation.

\subsection{Sorting Permutations by Rotations - Version 1}

We consider a permutation with $n$ elements. We want to sort (ascendingly) this permutation by performing only operations of the following type: (case $1$) we choose a position $i$ ($0\leq i\leq n$) and: all the numbers on the positions $1,\ldots,i$ are rotated one position to the left (or right, or we can choose the direction of the rotation), and all the numbers on the positions $i+1,\ldots,n$ are rotated one position to the right (or left, or we can choose the direction of the rotation); (case $2$) we choose a position $i$ ($0\leq i\leq n+1$) and all the numbers on the positions $1,\ldots,i-1$ are rotated $1$ position to the left (or right, or we can choose the direction of the rotation), and all the numbers on the positions $i+1,\ldots,n$ are rotated $1$ position to the right (or left, or we can choose the direction of the rotation); if $1\leq i\leq n$ then the element on the position $i$ is not moved.

We will first consider case $1$. We repeatedly choose position $n$, until element $1$ is on the first position of the permutation. Then, the problem is solved incrementally. At step $k$ ($2\leq k\leq n$), we have all the numbers from $1$ to $k-1$ on the positions $1,\ldots,k-1$. We now want to bring number $k$ on position $k$ (unless it's already there). In order to achieve this we will perform the following steps:\begin{enumerate}
\item We will repeatedly choose position $0$, until number $k$ is on the last position of the permutation.
\item Then, we repeatedly choose position $n-1$, until the numbers $1,\ldots,k-1$ are located on the positions $n-k+1,\ldots,n-1$ (at every rotation, number $k$ remains on the last position of the permutation).
\item Then, we repeatedly choose position $n$, until the numbers $1,\ldots,k$ are located on the positions $1,\ldots,k$.
\end{enumerate}

Case $2$ can be solved using a similar strategy. At the first step we rotate the entire permutation, until element $1$ is located on position $1$ (e.g. by repeatedly choosing $i=n+1$). Then, at each step $k$ ($2\leq k\leq n$), we choose position $i=0$ until element $k$ reaches position $n$. Afterwards, we repeatedly choose the position $i=n$, until the numbers $1,\ldots,k-1$ are located on the positions $n-k+1,\ldots,n-1$. Then, we repeatedly choose $i=n+1$, rotating the entire permutation until the numbers $1,\ldots,k$ are located on the positions $1,\ldots,k$.

In both cases we have to perform $O(n^2)$ operations. If we actually perform the rotations at every operation, the overall time complexity of the algorithm becomes $O(n^3)$. In order to maintain the $O(n^2)$ time complexity, we notice that we do not actually have to perform every operation. For instance, if, during the algorithm, we choose $p$ times consecutively the same position $i$, this is equivalent to rotating the two sides of the permutation by $p$ positions. Thus, when we want to move a number $k$ on the last position of the permutation we compute the number $p_1$ of operations which need to be performed ($p_1=pos(k)$ for rotations to the left, or $p_1=n-pos(k)$ for rotations to the right, where $pos(k)$ is the current position of element $k$ in the permutation). Then, when we need to move all the numbers $1,\ldots,k$ from the positions $q,q+1,\ldots,q+k-1$ of the permutation to the positions $r,r+1,\ldots,r+k-1$ ($1,\ldots,k$) of the permutation and they are located in a part containing the positions $1,\ldots,s$ ($s=n-1$ or $s=n$), we will compute the number $p_2$ of operations which need to be performed: $p_2=q-r$ (if $q\geq r$) or $p_2=q+s-r$ (if $q<r$) for rotations to the left; $p_2=r-q$ (if $r\geq q$) or $p_2=s-q+r$ (if $r<q$) for rotations to the right. Then, for each such values $p_1$ and $p_2$ we just perform a rotation by a multiple number of positions (in $O(n)$) time. This way, we only perform $O(1)$ rotations per step, obtaining an $O(n^2)$ time complexity.

\subsection{Sorting Permutations by Rotations - Version 2}

We consider a permutation of the numbers $1,\ldots,n$ again. We want to sort this permutation ascendingly, by using the following type of operations: we choose a position $p$ and: all the elements on the positions $1,\ldots,p-1$ are reversed, and all the elements on the positions $p+1,\ldots,n$ are also reversed. For instance, if the permutation is $Pe(1),\ldots,Pe(n)$ and we choose a position $p$, the new permutation will be: $Pe(p-1), Pe(p-2),\ldots,Pe(1),Pe(p),Pe(n),Pe(n-1),\ldots,Pe(p+1)$. We can also choose the positions $p=0$ and $p=n+1$. We will present a solution which performs $O(n)$ operations. At the first operation we will identify the position $p$ on which element $1$ is located. Then, we will choose the position $p+1$ (and we perform the operation). At the beginning of each step $k$ ($2\leq k\leq n$) the numbers $1,\ldots,k-1$ will be located on the positions $1,\ldots,k-1$ of the permutation, but in reverse order. The steps to be performed are the following:\begin{enumerate}
\item Let $p$ be the position on which element $k$ is located.
\item We perform an operation by choosing the position $p$. Now, the permutation contains the elements $1,\ldots,k$ on the positions $p-k+1,\ldots,p$, in increasing order.
\item After this, we will choose the position $p+1$, which will move the elements $1,\ldots,k$ on the first $k$ positions, in reverse order.
\item In the end, we perform one extra operation: we choose the position $0$.
\end{enumerate}

The algorithm performs $O(n)$ operations and the total time complexity is $O(n^2)$ ($O(n)$ per operation).

\subsection{Constrained Sorting of a Permutation by Swaps}

We consider a permutation of the numbers $1,\ldots,n$. We want to sort this permutation ascendingly by performing swaps. A call $Swap(p_i,p_j)$ swaps the elements on the positions $p_i$ and $p_j$ of the permutation. However, we can only call $Swap(p_i,p_j)$ for certain pairs of positions $(p_i,p_j)$.

A simple solution is the following. We construct a graph $G$ in which every vertex corresponds to a position of the permutation and we have an edge between vertices $i$ and $j$ if the call $Swap(i,j)$ is permitted. The presented algorithm runs in $n$ steps. At step $i$, all the numbers $1,\ldots,i-1$ are already on the positions $1,\ldots,i-1$ and we will bring element $i$ on position $i$ (without changing the positions of the elements $1,\ldots,i-1$). If element $i$ is not already on position $i$ at step $i$, then we perform the following actions:\begin{enumerate}
\item We search for a path in $G$ from the vertex corresponding to the current position $p$ of the element $i$ to the vertex corresponding to position $i$; let this path be $v_1=p, v_2, \ldots, v_k=i$
\item We perform, in this order, the calls: $Swap(v_1,v_2)$, $Swap(v_2,v_3),$ $\ldots,$ $Swap($ $v_{k-1},$ $v_k)$ (i.e. we call $Swap(v_j,v_{j+1})$ for $1\leq j\leq k-1$ in increasing order of $j$)
\item We perform the following calls: $Swap(v_j, v_{j-1})$ for $2\leq j\leq k-1$ in decreasing order of $j$
\end{enumerate}

We notice that after the swaps performed at step $2$, element $i$ arrives on position $i$, but many of the other elements may have been moved away from their positions. After performing step $3$ however, all the elements are brought back to their original positions (i.e. those before the swaps at step $2$), except for the element $q$ which was previously located on position $i$, and which will now be located on position $p$. Step $1$ can be implemented in $O(n+m)$ time (where $m$ is the number of edges of $G$), using any graph traversal algorithm (e.g. DFS or BFS). Steps $2$ and $3$ are trivially implemented in $k=O(n)$ time. Thus, the total time complexity is $O(n\cdot (n+m))$. If we initially split $G$ into connected components and maintain only a spanning of each connected component, then we can find a path between two positions $p$ and $i$ in $O(n)$ time (if it exists), because we will only consider the $O(n)$ edges of a spanning tree. Thus, the overall time complexity becomes $O(n^2)$.

\subsection{Minimum Cost Sorting of a Permutation by Swaps}

We consider a permutation $p(1),\ldots,p(n)$ of the numbers $1,\ldots,n$. We want to sort this permutation ascendingly by using swaps. Each number $i$ ($1\leq i\leq n$) has a cost $c(i)$. The cost of swapping two elements $x$ and $y$ in the permutation is $c(x)+c(y)$. We want to find a minimum cost strategy which sorts the permutation.

We will construct a graph with $n$ vertices, one for each number in the permutation, and $n$ directed edges: $(i,p(i))$ ($1\leq i\leq n$). This graph is a union of disjoint cycles. We will consider every cycle $Cy$ containing at least two vertices. For each vertex $i$ on the cycle, let $next(i)$ be the vertex following $i$ on $Cy$ and $prev(i)$ be the vertex preceeding $i$ on $Cy$. We have two options for bringing the values of the vertices on $Cy$ on their corresponding positions.

The first choice is the following:\begin{enumerate}
\item Let $q$ be the vertex of $Cy$ with the minimum cost $c(q)$.
\item We will repeatedly swap $q$ with every vertex on $Cy$, starting from $prev(q)$ and moving in reverse direction of the edges along $Cy$.
\item Let $SC$ be the sum of the costs of the vertices on $Cy$ and let $k$ be the number of vertices on $Cy$.
\item The cost of performing these swaps is $SC-c(q)+(k-1)\cdot c(q)$.
\end{enumerate}

The second choice is the following:\begin{enumerate}
\item Let $r$ be the element with the smallest value $c(r)$ in the entire permutation.
\item We swap the elements $q$ and $r$, and then we perform the same swaps as before, only using the element $r$ instead of $q$.
\item In the end, we swap $q$ and $r$ again (thus, $q$ also reaches its final position).
\item The cost of this choice is $SC-c(q)+(k-1)\cdot c(r)+2\cdot(c(q)+c(r))$ (where $SC$ and $k$ have the same meaning as before).
\end{enumerate}

For each cycle $Cy$ we will select the choice with the smallest cost. The time complexity of the algorithm is $O(n)$.

\subsection{Circular Sorting of a Multi-Permutation}

We consider a multi-permu-tation $p(1),\ldots,p(n)$, in which every number from $1$ to $k$ occurs at least once. We want to {\em circularly sort} the multi-permutation by performing swaps. The multi-permutation is circularly sorted if it is a circular permutation of the multi-permutation in which all the numbers of $p$ are sorted ascendingly. In order to sort the multi-permutation we can perform swaps. The cost of swapping the numbers on the positions $i$ and $j$ is $|i-j|$. We want to minimize the total number of swaps first and, if there are multiple strategies with the same number of swaps, we want to minimize the total cost of the performed swaps.

We will first use count-sort in order to construct in $O(n+k)$ time the ascending order of the values: $a(1),\ldots,a(n)$. Then, we will consider every circular permutation of $a(1),\ldots,a(n)$. Let $q(1),\ldots,q(n)$ be such a circular permutation. We will compute the best strategy (minimum number of swaps and minimum total cost) in order to sort the multi-permutation $p$ in the order given by $q$.

For every value $i$ ($1\leq i\leq k$) we will compute the list $Lp(i)$ of positions on which it occurs in $p$ and the list $Lq(i)$ of positions on which it occurs in $q$. The lists $Lp(i)$ and $Lq(i)$ are sorted ascendingly. Both sets of lists ($Lp(*)$ and $Lq(*)$) can be constructed in overall linear time, as follows: we traverse the multi-permutation $p$ ($q$) from the position $1$ to $n$ and we add each position $i$ at the end of $Lp(p(i))$ ($Lq(q(i))$) (all the lists are empty before the traversal). The lists $Lp(i)$ need only be computed once, in the beginning (before considering the first circular permutation $q$).

Then, we will initialize two variables: $ni=0$ and $ci=0$. For each value $i$ ($1\leq i\leq k$) we will merge the lists $Lp(i)$ and $Lq(i)$ into a list $Lr(i)$. Then, we will traverse the list $Lr(i)$ from the beginning to the end. Every time we find two consecutive equal elements $x$ in $Lr(i)$, we add $x$ to another list $Lc(i)$ (which is empty before we start traversing the list $Lr(i)$). Then, we will "merge" the lists $Lp(i)$ ($Lq(i)$) and $Lc(i)$ into a list $Lp'(i)$ ($Lq'(i)$). During the merge, as long as we haven't reached the end of any of the two lists, if the current element of $Lp(i)$ ($Lq(i)$) is equal to the current element of $Lc(i)$ then we do not add any of them to $Lp'(i)$ ($Lq'(i)$) and we simply move to the next elements; if they are different, then we add the current element of $Lp(i)$ ($Lq(i)$) to the end of $Lp'(i)$ ($Lq'(i)$). In the end, if any elements from $Lp(i)$ ($Lq(i)$) were not considered before considering all the elements of $Lc(i)$, we add these elements, in order, to the end of $Lp'(i)$ ($Lq'(i)$).

The lists $Lp'(i)$ and $Lq'(i)$ are sorted ascendingly and contain the positions $pos$ on which the element $i$ occurs in $p$ and, respectively, in $q$, but it does not occur on the same position in the other multi-permutation ($q$, respectively $p$). Let $Q(i,j)$ denote the $j^{th}$ element of a list $Q(i)$. We set $ni$ to the number of elements in $Lp'(i)$. Then, we traverse the elements of $Lp'(i)$ and we increase $ci$ by $|Lp'(i,j)-Lq'(i,j)|$ ($j$ is the current position of the element from $Lp(i)$; initially, $ci=0$). $ni$ is the minimum number of required swaps and $ci$ is the minimum total cost (for performing $ni$ swaps). If we were only interested in the minimum total cost, then we would use $Lp(i)$ ($Lq(i)$) instead of $Lp'(i)$ ($Lq'(i)$) when computing the cost $ci$.

We will maintain the pair $(ni,ci)$ with the minimum value of $ni$ (and, in case of ties, with the minimum value of $ci$). The time complexity is linear for each multi-permutation $q$. Since we consider $n$ multi-permutations $q$, the overall time complexity is $O(n^2)$.

\subsection{Sorting a Multi-Permutation by Swapping Adjacent Elements}

We consider two multi-permutations with $n$ elements: $p(1),\ldots,p(n)$ and $q(1),\ldots,q(n)$. Both $p$ and $q$ contain values from $1$ to $k$ and, moreover, they contain the same number of values $i$ ($1\leq i\leq k$), i.e. $q$ is obtained from $p$ by permuting its numbers somehow. We want to transform $p$ into $q$ by performing the following type of move: we can select a position $i$ ($1\leq i\leq n-1$) and swap the values $p(i)$ and $p(i+1)$. We want to compute the minimum number of swaps required to transform $p$ into $q$.

We will construct a permutation $r$ of the numbers $1,\ldots,n$, as follows. We first traverse the permutation $q$ from position $1$ to $n$ and, for every position $i$, we add $i$ at the end of a list $L(q(i))$ (we maintain $k$ lists $L(*)$ which are initially empty). Then, we initialize an array $idx(i)$ ($1\leq i\leq k$) to $0$ and we traverse the permutation $p$ from the first to the $n^{th}$ position. For every position $i$ ($1\leq i\leq n$) we: $(1)$ increment $idx(p(i))$ ; $(2)$ set $r(i)=L(p(i),idx(p(i)))$ (we denote by $L(u,v)$ the $v^{th}$ element of the list $L(u)$). The minimum number of swaps required to turn $p$ into $q$ is equal to the number of inversions of the permutation $r$.

We can compute the number of inversions of a permutation $r$ using several algorithms. The first algorithm consists of extending the merge-sort algorithm. We will maintain a global variable $ninv$ (which is initially $0$). Let's assume now that, during the merge-sort algorithm, we need to merge two sorted sequences, $A$ and $B$ (with $x$ and, respectively, $y$ elements), where $A$ consists of elements located to the left of the elements in $B$ within $p$. During the merging phase, we maintain two counters $i$ and $j$ (initialized at $1$), representing the current position in the sequences $A$ and, respectively, $B$. If $A(i)<B(j)$ we add $A(i)$ at the end of the sorted sequence and we increment $i$ by $1$; otherwise, we add $B(j)$ at the end of the sorted sequence, we increment $j$ by $1$ and we increment $ninv$ by $(x-i+1)$. When either $i$ exceeds $x$ or $j$ exceeds $y$, the remaining elements are added to the end of the sorted sequence. In the end, the number of inversions of $r$ is $ninv$. The time complexity of this approach is $O(n\cdot log(n))$.

A second algorithm consists of using a segment tree. Each of the $n$ leaves of the segment tree will have the values $1$. Inner nodes contain the sum of the values of the leaves in their subtrees. We will traverse the permutation $r$ from the position $1$ to position $n$. Like before, we will maintain the variable $ninv$ (initialized to $0$). For a position $i$, we query the segment tree in order to find the sum $S$ of the leaves from the interval $[1,r(i)-1]$. We will increment $ninv$ by $S$ and, after this, we set the value of the leaf $r(i)$ to $0$ (also updating the values of the leaf's ancestors). The time complexity of this approach is also $O(n\cdot log(n))$. If instead of the segment tree we use a block partition (in which every position from $1$ to $n$ has an initial value of $1$ and, after considering the position $i$, we set the value on the position $r(i)$ to $0$), we can obtain a time complexity of $O(n\cdot sqrt(n))$ (we denote by $sqrt(x)$ the square root of $x$).

The original problem can be solved with a linear ($O(n)$) time complexity if $k\leq 2$. We construct the lists of positions $Lp(1)$ and $Lq(1)$ on which the value $1$ occurs in $p$ and, respectively, $q$. These lists are sorted ascendingly and can be constructed in $O(n)$ time, as we discussed earlier. Then, the total number of required swaps is equal to the sum of the values $|Lp(1,i)-Lq(1,i)|$ (where $1\leq i\leq$ the total number of values equal to $1$ in $p$).

Note that in the case of the problem discussed in this section we can find the minimum number of swaps in $O(n\cdot log(n))$ time, but this number may be of the order $O(n^2)$. The strategy performing the swaps is quite straightforward. For every position $i$ from $1$ to $n$, if $p(i)\neq q(i)$ we will find the element $p(j)=q(i)$ on the smallest position $j>i$ and we swap it repeatedly with the element to its left, until it arrives on the position $i$.

\subsection{Sorting a Multi-Permutation by Grouping Identical Elements}

We consider a multi-permutation $p(1),\ldots,p(n)$, which contains every element $i$ ($1\leq i\leq k$) at least once. We want to rearrange the numbers in the multi-permutation such that all equal numbers are located on consecutive positions (i.e. grouped together in a contiguous sequence). In order to achieve this we can swap any pair of adjacent values (i.e. we can choose two positions $i$ and $i+1$ and swap the values $p(i)$ and $p(i+1)$; $1\leq i\leq n-1$). We want to find a strategy which performs the minimum number of swaps.

A first solution is the following. We will consider every possible permutation of the $k$ distinct values (there are $k!$ such permutations). Then, let this permutation be $q(1),\ldots,q(k)$. We will compute the minimum number of swaps required to bring all the values equal to $q(1)$ first, then all the values equal to $q(2)$, and so on.

Before considering any permutation, we will compute the values $cnt(*)$, where $cnt(j)$=the number of elements equal to $j$ in the multi-permutation $p$ (we have $cnt(1)+\ldots+cnt(k)=n$). Then, for a given permutation $q(1),\ldots,q(k)$, we will compute the values $scnt(q(i))=cnt(q(1))+\ldots+cnt(q(i-1))$ ($scnt(q(1))=0$ and $scnt(q(2\leq i\leq k))=scnt(q(i-1))+cnt(q(i-1))$ (these values are computed in the order given by the permutation $q$). Then, we initialize a set of values $idx(1\leq i\leq k)$ to $0$ and we traverse the multi-permutation $p$ (from $i=1$ to $n$). During the traversal we will construct a permutation $r$. For each position $i$ of $p$, we will first increment $idx(p(i))$ by $1$ and then we set $r(i)=scnt(p(i))+idx(p(i))$. In the end, $r$ is a permutation and the minimum number of swaps required to sort the multi-permutation $p$ according to the constraints of the permutation $q$ is equal to the number of inversions of $r$. Computing the number of inversions of a permutation with $n$ elements can be performed in $O(n\cdot log(n))$ time, as was shown in the previous subsection. Thus, we obtained an algorithm with a time complexity of $O(k!\cdot n\cdot log(n))$.

We can improve our solution as follows. In the beginning we will compute a $k-by-k$ matrix $num$, where $num(a,b)$=the number of pairs of positions $(i,j)$, such that $p(i)=a$, $p(j)=b$, and $i<j$. This matrix can be easily computed in $O(n^2)$ time. We simply initialize it to all zeroes and then we consider every pair of positions $(i,j)$ ($1\leq i<j\leq n$) and increment $num(p(i),p(j))$ by $1$. However, we can compute it more efficiently. We initialize the matrix to zero and then we traverse the multi-permutation from the position $1$ to $n$. We will maintain an array $cnt$, where $cnt(j)$=the number of values equal to $j$ encountered so far. Initially, $cnt(1\leq j\leq k)=0$. For every position $i$ ($1\leq i\leq n$) we will consider every value $j$ ($1\leq j\leq k$) and we will increment $num(j,p(i))$ by $cnt(j)$; after this, we increment $cnt(p(i))$ by $1$. Thus, we computed the $num$ matrix in $O(n\cdot k)$ time.

After computing the matrix $num(*,*)$ we will consider again all the possible $k!$ permutations $q(1),\ldots,q(k)$ of the distinct values of $p$. For a given permutation $q(1),\ldots,q(k)$, the minimum number of swaps required to sort the multi-permutation $p$ according to the constraints imposed by the permutation $q$ is equal to the sum of all the values $num(q(j),q(i))$ with $1\leq i<j\leq k$. Thus, we obtained a solution with an $O(n\cdot k+k!\cdot k^2)$ time complexity. If we generate the $k!$ permutations in the Steinhaus-Johnson-Trotter order (also called {\em transposition order}), then two consecutive permutations differ from each other in exactly two consecutive positions $i$ and $i+1$. Thus, for the first generated permutation we use the algorithm described above. Then, when we generate a new permutation $q(1),\ldots,q(k)$ which differs from the previously generated permutation on the positions $i$ and $i+1$, we will compute the number of swaps as follows. Let $V$ be the number of swaps for the previous permutation. The number of swaps $V'$ for the current permutation will be equal to $V'=V+num(q(i+1),q(i))-num(q(i),q(i+1))$. Thus, the time complexity is now $O((n+k)\cdot k+k!)$ (if we generate every new permutation in $O(1)$ (amortized) time).

Another solution, also based on computing the matrix $num(*,*)$ is to use dynamic programming. We will compute the values $nmin(S)$=the minimum number of swaps required to bring the values belonging to the set $S$ in some order before all the other values of the multi-permutation (and ignoring those values which are not part of $S$). $S$ is a subset of $\{1,\ldots,k\}$. We will consider the subsets $S$ in increasing order of their number of elements (or in lexicographic order). We have $nmin(\{\})=0$. For $|S|>0$ we proceed as follows. For every element $i$ from $S$ we will consider the case when the values equal to $i$ are placed last (after all the other values in $S$). We will compute $nmin'(S,i)=nmin(S\setminus \{i\})$ plus the sum of the values $num(i,j)$, where $j\in S$ and $j\neq i$. We have $nmin(S)=min\{nmin'(S,i)|i\in S\}$. The final result is $nmin(\{1,\ldots,k\})$. The time complexity of this solutions is $O(n\cdot k+2^k\cdot k^2)$.

This time complexity can be slightly improved as follows. Initially, we will compute the values $Sum(S,i)$ for every subset $S$ and every element $i\in S$, representing the sum of the values $num(i,j)$ for $j\in S$ and $j\neq i$. We will compute these values in increasing order of the elements of $S$ (or in lexicographic order of the subsets $S$). We have $Sum(\{i\},i)=0$. For $|S|\geq 2$ let $j\neq i$ be an arbitrary element of $S$. We have $Sum(S,i)=Sum(S\setminus \{j\},i)+num(i,j)$. This takes $O(k\cdot 2^k)$ time overall. Thus, in the algorithm from the previous paragraph we can compute $nmin'(S,i)$ in $O(1)$ time instead of $O(k)$ by using the value $Sum(S,i)$.

\section{Set Maintenance based on the Union-Find and Split-Find Problems}

The Union-Find problem consists of supporting efficiently the following two operations on a family of disjoint sets: {\em Union(A,B)} performs the set union of the sets identified by $A$ and $B$; {\em Find(x)} returns the identifier of the set which contains the element $x$. The Union-Find problem has been studied extensively, due to its wide range of applications. In this section we propose several simple extensions, which, nevertheless, are important from a practical point of view.

We consider the elements of each set arranged in a row. At first, we have $n$ elements, identified with numbers from $1$ to $n$; each element $i$ forms a set on its own and has a weight $w(i)$. We consider a sequence of two types of operations: {\em union} and {\em query}. A union specifies the identifiers of two elements $x$ and $y$ and a direction $d$ ({\em left} or {\em right}). Let $rx=Find(x)$ be the identifier of the set containing $x$ and $ry=Find(y)$ be the identifier of the set containing $y$. The union operation combines the two sets into a single set, in the following way. If $d=left$, then the elements of the set $rx$ are placed to the left of those in the set $ry$; otherwise, the elements of the set $rx$ are placed to the right of those in the set $ry$. A query specifies an element $x$ and asks for the the aggregate weight of the elements $y$ in the same set as $x$ which are located (strictly) to the left of $x$ (using a pre-specified aggregation function $aggf$).

We will represent the sets as rooted trees. The root of each tree will be the representative element (the identifier) of the set. For each element $x$ we store its parent in the tree ($parent(x)$), a value $wp(x)$ and the aggregate weight of the elements in its subtree ($wagg(x)$). We first present a solution for the case when $aggf$ has an inverse $aggf^{-1}$. We initialize $parent(x)$ to $null$, $wp(x)$ to the neutral element of $aggf$ (e.g. $0$, for $aggf=+$, $xor$; $1$ for $aggf=*$) and $wagg(x)$ to $w(x)$ for each element $x$.

For a query operation with the argument $x$, the answer will be the aggregate of the $wp(y)$ values, where $y$ is an ancestor of $x$ in its tree (including $x$).

For a union operation, we compute the representatives of the sets containing $x$ and $y$, $rx$ and $ry$, by following the parent pointers all the way up to the tree roots. We will use the {\em union by rank} heuristic. If the height (or total number of nodes) of the tree rooted at $rx$ is smaller than or equal to that of the tree rooted at $ry$, we set $parent(rx)=ry$; otherwise, we set $parent(ry)=rx$. After setting the parent pointer of one of the two representatives, we consider the direction $d$ of the union:\begin{itemize}
\item If $d=left$ and $parent(rx)=ry$, we set $wp(ry)=wp(ry)$ $aggf$ $wagg(rx)$ and, after this, $wp(rx)=wp(rx)$ $aggf$ $(wp(ry))^{-1}$.
\item If $d=left$ and $parent(ry)=rx$, then we set $wp(ry)=(wp(ry)$ $aggf$ $wagg(rx))$ $aggf$ $(wp(rx))^{-1}$.
\item If $d=right$ and $parent(rx)=ry$, then we set $wp(rx)=(wp(rx)$ $aggf$ $wagg(ry))$ $aggf$ $(wp(ry))^{-1}$.
\item If $d=right$ and $parent(ry)=rx$, then we first set $wp(rx)=wp(rx)$ $aggf$ $wagg(ry)$; then, we set $wp(ry)=wp(ry)$ $aggf$ $(wp(rx))^{-1}$ (considering the updated value of $wp(rx)$).
\end{itemize}

In the end, if $parent(rx)=ry$ then we set $wagg(ry)=wagg(ry)$ $aggf$ $wagg(rx)$; otherwise, we set $wagg(rx)=wagg(rx)$ $aggf$ $wagg(ry)$.

Since we use the union by rank heuristic, the height of every tree is $O(log(n))$. Thus, performing a query takes $O(log(n))$ time (because an element has $O(log(n))$ ancestors). We can improve the query time by also using the {\em path compression} heuristic. The path compression heuristic works as follows. Every time we need to traverse all the ancestors of an element $x$ (from $x$ towards the root of its tree), we change the tree and make $parent(y)=rx$, where $rx$ is the root of element $x$'s tree and $y$ is on the path between $x$ and $rx$ (including $x$ and excluding $rx$); when doing this, we need to take care of also changing the values $wp(y)$. Let's assume that the path from $x$ to $rx$ consists of the elements: $v_{1}=x$, $v_{2}=parent(x)$, $v_{3}=parent(parent(x))$, $\ldots$, $v_{h}=rx$ (the root). During a {\em Find(x)} call or a query with argument $x$, we compute $wpagg(i)=wp(v_{i})$ $aggf$ $\ldots$ $aggf$ $wp(v_{h-1})$ ($1\leq i\leq h-1$) ($wpagg(1\leq i\leq h-2)=wp(v_{i})$ $aggf$ $wpagg(i+1)$ and $wpagg(h-1)=wp(v_{h-1})$). We set $parent(v_{i})$ to $rx$ and we set $wp(v_{i})$ to $wpagg(i)$ ($1\leq i\leq h-1$). The overall (amortized) time complexity becomes $O(n\cdot \alpha (m,n)$), where $m(\geq n)$ is the total number of operations and $\alpha (m,n)$ is the inverse of the Ackermann function.

We can also use the path compression technique without the complex union rules (and/or heuristics), for the general case where $aggf$ is only commutative and associative (e.g. $aggf=max$, $min$). We will maintain the $wagg(x)$ values as before. Each edge $(u,parent(u))$ of a tree will have a value $wskip(u)$. When we unite two sets with representatives $rx$ and $ry$ such that the set $rx$ will be located to the left of the set $ry$, we set $parent(ry)=rx$ and set $wskip(ry)=wagg(rx)$. Afterwards, we update $wagg(rx)$ (we set it to $wagg(rx)$ $aggf$ $wagg(ry)$). The aggregate weight of the elements located strictly to the left of an element $x$ is the aggregate of the $wskip$ values of the edges on the path from $x$ to its set representative (tree root). At every $Find(x)$ (or query with $x$ as an argument) operation, we compute the set representative $rx$ and then the values $wskipagg(y)=wskip(y)$ $aggf$ $wskipagg(parent(y))$ (where $y$ is on the path from $x$ to $rx$, including $x$; $wskipagg(rx)=undefined$, i.e. it is the neutral element of the aggregation function). Then, we set $parent(y)=rx$ for every node $y$ on the path from $x$ to $rx$ (including $x$, if $x\neq rx$, and excluding $rx$), as well as $wskip(y)=wskipagg(y)$.

The solution presented so far for this problem considered the online case (i.e. every query and union operation was handled as soon as it was received). In the offline case we can construct a somewhat simpler solution. We will first process all the union operations (in order), ignoring the queries. Like before, we will maintain disjoint sets with a tree structure. For each set with its representative $rx$ we will maintain $leftmost(rx)$ and $rightmost(rx)$, representing the index of the leftmost and rightmost elements in the set. Initially, we have $leftmost(x)=rightmost(x)=x$ for every element $x$. When we perform a union and the set identified by $rx$ is placed to the left (right) of the set identified by $ry$, we will add a directed edge from $rightmost(rx)$ to $leftmost(ry)$ (from $rightmost(ry)$ to $leftmost(rx)$). After this, if we need to set $parent(rx)=ry$ ($parent(ry)=rx$), then the new root $r$ will be $ry$ ($rx$). If $rx$ was placed to the left of $ry$ then we will have $leftmost(r)=leftmost(rx)$ and $rightmost(r)=rightmost(ry)$; otherwise, we will have $leftmost(r)=leftmost(ry)$ and $rightmost(r)=rightmost(rx)$.

After processing all the unions, we consider the graph composed of the $n$ elements as vertices and the added directed edges. From every vertex there is at most one outgoing edge. Thus, the graph is the union of a set of disjoint chains (directed paths). We will arrange all the elements consecutivey in the order in which they appear on their paths. Then we will concatenate the orderings corresponding to each path, considering an arbitrary order of the paths. Thus, we obtain a permutation $p(1),\ldots,p(n)$, such that: if $p(i)$ is not the rightmost element in its set, then $p(i+1)$ is the element from its set which is immediately to its right. Then, we will construct a 1D data structure $DS$ over the ordering of these elements which will allow us to answer range queries efficiently (a query consists of the aggregate weight of the elements $p(i),\ldots,p(j)$ whose positions are contained in a given range $[i,j]$). If the $aggf$ function is invertible, we can compute prefix "sums" in order to answer a query in $O(1)$ time. If $aggf$ is {\em min} or {\em max} we can preprocess the elements in order to answer range minimum (maximum) queries in $O(1)$ time. Otherwise, we can construct a segment tree over the $n$ elements which will allow us to answer aggregate queries in $O(log(n))$ time. Moreover, for each element $i$ ($1\leq i\leq n$) we will store its position $pos(i)$ in this ordering (i.e. $p(pos(i))=i$). Then, we will process the entire sequence of operations from the beginning. We reinitialize the disjoint sets and, like before, we will maintain the $leftmost(*)$ and $rightmost(*)$ values. This time we will also process the queries. In the case of a query for an element $x$, we first compute $rx$ the representative of the set containing $x$. The answer to the query is obtained by range querying $DS$ with the range $[pos(leftmost(rx)),pos(x)-1]$.

We will now present an extension of the Split-Find problem, which was brought to our attention by R. Berinde. There are $n$ elements placed consecutively in a row (from $1$ to $n$). Initially, they are all part of the same set (interval). We can perform two types of operations. The operations may split an interval into two intervals or undo a split (unite two intervals back into a larger interval). The only initial interval $[1,n]$ has color $C$. The $Split(i, k, C_{left}, C_{right})$ operation considers the interval $[i,j]$ starting at $i$ and a position $k$ ($i\leq k<j$). The interval $[i,j]$ is split into the intervals $[i,k]$ and $[k+1,j]$. The interval $[i,k]$ is colored with color $C_{left}$ and the interval $[k+1,j]$ is colored using color $C_{right}$. The $Undo(k)$ operation considers a position $k$ where an interval $[i,j]$ was previously split and unites the two intervals $[i,k]$ and $[k+1,j]$, thus forming the interval $[i,j]$ back. The interval $[i,j]$ will get the color it had before the split. Obviously, this operation can only be used if the intervals obtained after the corresponding {\em Split} exist (i.e. they have not been split further or, if they have, they were put back together). A third operation $Query(i)$ asks for the color of the interval starting at position $i$ (if such an interval exists).

We will present here a solution which takes $O(1)$ time per operation, no matter how the operations are mixed into the sequence of operations, and uses $O(n)$ memory. We will maintain several arrays: $start$, where $start(i)=1$ if an interval starts at position $i$ (and $0$, otherwise); $col$, where $col(i)$=the color of an interval starting at position $i$; $isplit$, $jsplit$, $csplit$, where the interval $[isplit(k), jsplit(k)]$ was the last interval split at the position $k$ and $csplit(k)$=the color of the interval $[isplit(k),jsplit(k)]$ before the last split which had the position $k$ as a split parameter; $jj$, where $jj(i)$=the finish endpoint of the interval starting at $i$ (it makes sense only if $start(i)=1$). Initially, we have $start(1)=1$, $start(i>1)=0$, $jj(1)=n$ and $col(1)=C$.

A $Split(i,k,C_{left},C_{right})$ operation performs the following actions (assuming $start(i)=1$):\begin{enumerate}
\item $j=jj(i)$
\item $isplit(k)=i$
\item $jsplit(k)=j$
\item $csplit(k)=col(i)$
\item $start(k+1)=1$
\item $col(i)=C_{left}$
\item $col(k+1)=C_{right}$
\item $jj(i)=k$
\item $jj(k+1)=j$
\end{enumerate}

A $Query(i)$ operation returns $col(i)$, if $start(i)=1$, or $undefined$, otherwise. $Undo(k)$ performs three steps:\begin{enumerate}
\item $start(k+1)=0$
\item $col(isplit(k))=csplit(k)$
\item $jj(isplit(k))=jsplit(k)$
\end{enumerate}

We can also support an operation $Undo(k,C')$, having the same meaning as $Undo(k)$, except that the interval which is formed back will get the color $C'$ instead of the color it had before the split. $Undo(k,C')$ contains the same steps as $Undo(k)$, except that step $2$ is: $col(isplit(k))=C'$. We can also define the operations $Undo'(i)$ and $Undo'(i,C')$, where $i$ is the beginning of an interval which will be united back with the interval after it. $Undo'(i[,C'])$ is equivalent to calling $Undo(jj(i)[,C'])$.

\section{Minimum Cost Activation and Deactivation of Resources in a Graph with Bounded Treewidth}

We consider an undirected graph $G$ with $p$ vertices and $m$ edges, together with a tree decomposition of $G$, whose (tree)width is bounded by a small constant $tw$. The tree decomposition $T$ contains $n$ nodes ($n=O(p)$). Every node $X$ contains a subset $S(X)$ of vertices of $G$. The subsets of vertices of the nodes of any tree decomposition have the following properties:\begin{itemize}
\item if a vertex $u$ of $G$ belongs to both $S(X)$ and $S(Y)$ then $u$ belongs to the subsets $S(Z)$ of every node $Z$ on the path between $X$ and $Y$ in $T$
\item for every edge $(u,v)$ of $G$ there exists at least one subset $S(X)$ such that both $u$ and $v$ belong to $S(X)$
\item the size of every subset $S(X)$ is at most $tw$
\end{itemize}

Every vertex $u$ of $G$ has an associated resource which can be in one of the following two states: $active$ or $inactive$. The initial state of the resource at a vertex $u$ is $I(u)$ ($1$ for $active$, or $0$ for $inactive$). The final desired state of the resource at a vertex $u$ is $F(u)$. In order to change the states of the resources, we can repeatedly perform the following action: we can select a vertex $u$ of $G$ and change the state of the resource at $u$, as well as the states of the resources of all the neighbors $v$ of $u$. Changing the state of a resource means bringing it into the state opposite from the current one (i.e. from active to inactive, or from inactive to active). Selecting a vertex $u$ for performing the action incurs a cost $C(u)\geq 0$. We want to find a strategy which brings every resource into its final state and which incurs a minimum total cost.

Solutions for general graphs (without bounded treewidth), as well as for several particular graphs have been discussed in \cite{gauss}. However, none of those solutions can match the time complexity of the algorithm we will present in this section.

The first observation is that we never need to select a vertex $u$ more than once. Thus, a vertex $u$ is {\em selected} if it was selected once, and {\em not selected} otherwise.

We will start by presenting a solution for the case in which $G$ is a tree. In this case we do not need to use the tree decomposition $T$. We will choose an arbitrary vertex $r$ as the root of $G$, thus defining parent-son relationships. For every vertex $u$ from $G$ we will compute the values $Cmin(u,state,sel)$=the minimum total cost for bringing all the resources within vertex $u$'s subtree to their final states (except possibly for the resource at vertex $u$), such that (the resource at) vertex $u$ is in the state {\em state} and vertex $u$ has been selected (if $sel=1$) or not (if $sel=0$). These values will be computed bottom-up.

For a leaf vertex $u$ we have $Cmin(u,I(u),0)=0$, $Cmin(u,1-I(u),1)=C(u)$ and $Cmin(u,I(u),1)=Cmin(u,1-I(u),0)=+\infty$. For an inner vertex $u$ we will compute the required values as follows. Let $ns(u)$ be the number of sons of the vertex $u$ and let $s(u,j)$ be the $j^{th}$ son of the vertex $u$, in some arbitrary order ($1\leq j\leq ns(u)$). We will first consider the cases with $sel=0$. We will compute the values $Sum(u,k)$ ($k=1,2$) as the sum of the values $min\{Cmin(s(u,j),(F(s(u,j))+k)$ $mod$ $2,0),Cmin(s(u,j),(F(s(u,j))+k)$ $mod$ $2,1)\}$ with $1\leq j\leq ns(u)$. We also compute $NumSel(u,k)$ as the number of sons $s(u,j)$ for which $min\{Cmin(s(u,j),(F(s(u,j))+k)$ $mod$ $2,q)|q=0,1\}=Cmin(s(u,j),(F(s(u,j))+k)$ $mod$ $2,1)$ ($1\leq j\leq ns(u)$). $NumSel(u,k)$ is the number of selected sons which contribute to the sum $Sum(u,k)$. Let $DifMin(u,k)=$ $min\{|Cmin(s(u,j),$ $(F(s(u,j))+k)$ $mod$ $2, 1)-Cmin(s(u,j),$ $(F(s(u,j))+k)$ $mod$ $2, 0)||1\leq j\leq ns(u)\}$.

We have $Cmin(u,(I(u)+NumSel(u,0))$ $mod$ $2,0)=Sum(u,0)$ and $Cmin(u,$ $(I(u)+NumSel(u,0)+1)$ $mod$ $2,0)=Sum(u,0)+DifMin(u,0)$.

For the case $sel=1$ we have $Cmin(u,(I(u)+NumSel(u,1)+1)$ $mod$ $2, 1)=Sum(u,1)+C(u)$ and $Cmin(u,(I(u)+NumSel(u,1))$ $mod$ $2,1)=Sum(u,1)+DifMin(u,1)+C(u)$. 

The minimum total cost for bringing every resource into its final state is $min\{$ $Cmin(r,F(r),q)|q=0,1\}$. The time complexity of the presented algorithm is $O(p)$ (i.e. linear in the number of vertices of $G$).

We will now return to our original problem. For every node $X$ of $T$, let $num(X)$ be the number of vertices $u\in S(X)$ and let these vertices be $u(X,1),\ldots,u(X,num(X))$. We will compute the values $Cmin(X,$ $(state(1),\ldots,state(num(X))),$ $(sel(1),$ $\ldots,$ $sel(num(X))))$, representing the minimum total cost for bringing to their final states the resources of all the vertices of $G$ belonging to subsets $S(Y)$ where $Y$ is in node $X$'s subtree (except possibly for the vertices $u(X,j)$, $1\leq j\leq num(X)$), such that the resource in every vertex $u(X,j)$ has changed its state an even (if $state(j)=0$) or odd (if $state(j)=1$) number of times, and the vertex $u(X,j)$ has been selected (if $sel(j)=1$) or not (if $sel(j)=0$) ($1\leq j\leq num(X)$).

For every node $X$ and every combination $CSel=(CSel(1),\ldots,CSel(num(X)))$ we will compute $NumSel(X,i,CSel)$ ($1\leq i\leq num(X)$) as the number of vertices $u(X,j)\in S(X)$ ($1\leq j\leq num(X)$) with $CSel(j)=1$ and which are neighbors with the vertex $u(X,i)$, and $SumCSel(X,CSel)$ as the sum of the costs $C(u(X,j))$ of the vertices $u(X,j)$ with $CSel(j)=1$ ($1\leq j\leq num(X)$).

For each node $X$ and each son $Y$ of $X$, we define the set $Common(X,Y)$, containing those vertices belonging to the intersection of $S(X)$ and $S(Y)$: $v(Y,1),\ldots,$ $v(Y,NumCommon(Y))$, where $NumCommon(Y)$ is the number of vertices in the set $Common(X,Y)$. If $r$ is the root of $T$ then we define $Common(parent(r),r)=\{\}$.

For every node $X$ and for every combination $CSel'=(CSel'(1)$, $\ldots,$ $CSel'($ $NumCommon(X)))$ we will compute $NumSelCommon(X,i,CSel')$ as the number of vertices $v(X,j)\in Common(parent(X),X)$ ($1\leq j\leq NumCommon(X)$) for which $CSel'(j)=1$ and which are neighbors with $v(X,i)$. We will also compute $SumCSelCommon(X,CSel')$ as the sum of the costs $C(v(X,j))$ of the vertices $v(X,j)$ $\in Common(parent(X),X)$ ($1\leq j\leq NumCommon(X)$) for which $CSel'(j)=1$.

We will also maintain two hash tables at every node $X$ of $T$. The first one will map every vertex $u(X,j)$ to its corresponding index in the set $Common(parent(X),X)$ (i.e. based on this hash table we will be able to find out if $u(X,j)\in Common(parent(X),$ $X)$ and, if so, we will be able to find the index $p$ associated to $u(X,j)$ such that $u(X,j)=v(X,p)$). The second hash table will map every vertex $v(X,j)$ to its corresponding index in the set $S(X)$ (i.e. based on this hash table we will be able to find the index $q$ associated to $v(X,j)$ such that $v(X,j)=u(X,q)$). Every operation on each of the hash tables takes constant time.

Let's assume first that we computed all the values $Cmin(X,(*,\ldots,*),(*,\ldots,*))$ for a node $X$ of $T$. After having these values computed, we will compute the values $CminCommon(X,$ $CState',$ $CSel')$ = $min\{$ $Cmin(X,$ $CState,$ $CSel)$ $|$ $CSel'(j)=CSel(q)$ and $CState'(j)=CState(q)$ (such that $v(X,j)=u(X,q)$) for every $1\leq j\leq NumCommon(X)$, and $CSel(j')=0$ or $1$ for every vertex $u(X,j')\notin Common($ $parent(X),$ $X)$ ($1$ $\leq$ $j'$ $\leq$ $num(X)$), and $CState(j'')=$ $((F(u(X,j''))+$ $I(u(X,j'')))$ $mod$ $2$) for every vertex $u(X,j'')$ such that $u(X,j'')\in S(X)$ and $u(X,j'')\notin Common($ $parent(X),X)$ ($1\leq j''\leq num(X)$) $\}$. The easiest way to perform these computations is to first initialize $CminCommon(X,$ $(*,\ldots,*),$ $(*,\ldots,*))=+\infty$. Then, we will consider every possible pair of combinations $(CState,CSel)$. We extract $CState'$ from $CState$ and $CSel'$ from $CSel$ (by maintaining only those indices $j$ for which $u(X,j)\in Common(parent(X),X)$ and reordering the values corresponding to those indices in the order corresponding to the indices of the vertices from the set $Common(parent(X),X)$ (the $q^{th}$ component of $CState'$ and $CSel'$ corresponds to $v(X,q)$ ($1\leq q\leq NumCommon(X)$)). Then, we set $CminCommon(X,$ $CState',$ $CSel')=min\{CminCommon(X,$ $CState',$ $CSel'),$ $Cmin(X,CState,CSel)\}$.

We will now show how to compute the values $Cmin(X,(*,\ldots,*),(*,\ldots,*))$ of every node $X$, when traversing the tree $T$ bottom-up (from the leaves towards the root). For every node $X$ we will first initialize $Cmin(X,(*,\ldots,*),(*,\ldots,*))=+\infty$. Then, we will consider every possible combination $CSel$ and we will compute the values $Cmin(X,(*,\ldots,*),CSel)$.

If $X$ is a leaf in $T$, then the state of the resource from a vertex $u(X,j)$ is $(I(u(X,j))+NumSel(X,j,CSel)+CSel(j))$ $mod$ $2$. Thus, we will set $Cmin(X,$ $CState=(CState(i)=(NumSel(X,i,CSel)+CSel(i))$ $mod$ $2|1\leq i\leq num(X)),$ $CSel)=SumCSel(X,CSel)$.

If the node $X$ is not a leaf in $T$, then let $Y(1),\ldots,Y(ns(X))$ be the $ns(X)$ nodes which are the sons of $X$ (in an arbitrary order). We will compute the values $Cmin'(X,j,CState,CSel)$=the minimum total cost of bringing into their final states the resources of all the vertices of $G$ located in the nodes $Z$, where either $Z=X$ or $Z$ is a descendant of one of the nodes $Y(q)$ (including $Y(q)$) ($1\leq q\leq j$), except possibly for the resources of the vertices of $S(X)$, whose states were changed a number of times whose parity is defined by $CState$. $CSel$ is the combination we fixed earlier.

For every combination $CState=(CState(i)=(NumSel(X,i,CSel)+CSel(i))$ $mod$ $2|1\leq i\leq num(X))$, we will set $Cmin'(X,0,CState,CSel)=SumCSel(X,$ $CSel)$; for the other possible combinations $CState$ we will set $Cmin'(X,0,CState,$ $CSel)=+\infty$. After this, we will consider the sons $Y(j)$ of $X$, in increasing order of $j$. We will first initialize all the values $Cmin'(X,j,(*,\ldots,*),CSel)=+\infty$. Then, we will construct a new combination $CSel'$ obtained by removing from $CSel$ the components $q'$ corresponding to those vertices $u(X,q')$ which do not belong to $Common(X,Y(j))$. Then, the remaining components are ordered such that the $q^{th}$ component of $CSel'$ refers to $v(Y,q)$ ($1$ $\leq$ $q$ $\leq$ $NumCommon(Y(j))$). 

After this, we will consider all the $2^{NumCommon(Y(j))}$ possible combinations (tuples) $CState'$ {\em =} $(CState'(1),$ $\ldots,$ $CState'(NumCommon(Y(j))))$. For each combination $CState'$ we will consider every possible combination $CState=(CState(1),$ $\ldots,$ $CState(num(X)))$. For each such combination $CState$ we will construct a new combination $CState''$, where: $CState''(i)=$ $CState(i)$ if $u(X,$ $i)$ $\notin$ $Common(X,$ $Y(j))$, and $CState''(i)=((CState(i)+CState'(q)+NumSelCommon(Y(j),q,CSel'))$ $mod$ $2$) if $u(X,i)\in$ $Common(X,Y(j))$ and $u(X,i)=v(Y(j),q)$ ($1\leq i\leq num(X)$). Then, we will set $Cmin'(X,j,CState'',CSel)=min\{Cmin'(X,j,CState'',CSel), Cmin'(X,$ $j-1,CState,CSel)+CminCommon(Y(j),CState',CSel')$ $-$ $SumCSelCommon(Y($ $j),$ $CSel')\}$.

In the end, we will have $Cmin(X,CState,CSel)=Cmin'(X,ns(X),CState,$ $CSel)$ (for every possible combination $CState$).

The minimum cost we are looking for is $min\{Cmin(r,CState=(CState(i)=(I(u(r,i))+F(u(r,i)))$ $mod$ $2|1\leq i\leq num(r)),*)\}$, where $r$ is the root node of $T$. With a careful implementation we can obtain an $O(n\cdot tw\cdot 2^{3\cdot tw})$. If we precompute all the transformations for each (generic) pair $t_1(CSel,$ {\em subset of indices}$)\rightarrow CSel'$ and $t_2(CState,$ {\em subset of indices}$)\rightarrow CState'$, and we establish a consistent ordering for the vertices in each subset $S(X)$ and $Common(U,V)$ (e.g. the vertices are ordered increasingly according to their identfiers), then we can obtain a time complexity of $O(n\cdot 2^{3\cdot tw})$.

\section{Related Work}

Data transfer optimization problems have been considered in many papers, because of their highly important practical applications. References \cite{bdts} and \cite{concsched} present offline and online algorithms for several multicriteria data transfer optimization problems (e.g. deadline-constrained data transfer scheduling). \cite{opt_ched_two_comm_flows} considers the optimal scheduling of two communication flows on multiple disjoint paths in order to minimize the makespan. Applications of several data structures to resource reservations were discussed in \cite{centr}. The technique for computing optimal average subsets of edges (or vertices) has been mentioned (in a similar form) in \cite{eppstein_avg}.

String and permutation sorting problems have also been considered from multiple perspectives and considering various constraints: \cite{string_sort} considers the sorting of sequences by interchange operations, while \cite{reversals} considers sorting permutations by reversal operations.

A permutation sorting problem related to the ones mentioned in this paper was proposed as a task at the Baltic Olympiad in Informatics 2007: We have a permutation $p(1),\ldots,p(n)$ of the numbers $1,\ldots,n$. We want to sort this permutation in ascending order by using a sequence of the operations $Move(i,j)$ (which removes the element from position $i$ in the permutation and inserts it at position $j$). The cost of an operation $Move(i,j)$ is $i+j$ and we are interested in sorting the permutation with a minimum total cost. The solution to this problem is based on two important observations: $1)$ every element is moved at most once; $2)$ the moved elements are moved in decreasing order of their values. These observations lead to a dynamic programming solution. We compute $Cmin(i,j)$=the minimum total cost of moving the elements $i,\ldots,n$ such that they are in their correct relative order and element $i$ is located at position $j$. We have $Cmin(n+1,n+1)=0$ and $Cmin(n+1,1\leq j\leq n)=+\infty$. We will compute the values $Cmin(i,*)$ in decreasing order of $i$. We will always have the possibility of moving the element $i$ or of not moving it. However, when not moving element $i$, we will consider a compensation cost, such that elements smaller than $i$ may consider that $i$ was moved. We will start by computing the array $pos(k)$=the position on which element $k$ is located in the original permutation ($1\leq k\leq n$). Now, before computing the values $Cmin(i,*)$, we will first compute the value $pos'(i)$, which is equal to $1$ plus the number of values $k$ such that $1\leq k\leq i-1$ and $pos(k)<pos(i)$ (because these are the only elements which are still to the left of element $i$ in the permutation). We will consider that element $i$ is moved right before the element $i+1$. Thus, we will consider all the values $j$ from $1$ to $n+1$, in increasing order. While traversing these values, we will maintain a variable $pdest$, which is initially equal to $1$. When we reach a value $j$, we will first update $pdest$: if $p(j)<i$ then $pdest=pdest+1$ (we consider $p(n+1)=n+1$). Then, we set $Cmin(i,j)=Cmin(i+1,j)+pos'(i)+pdest$. We will now consider the case when $i$ is not moved. We will consider all the positions $j$ from $pos(i)+1$ up to $n+1$ in increasing order; during this time, we will maintain a variable $extra\_cost$ (which is initially $0$). When we reach such a position $j$, we set $Cmin(i,pos(i))=min\{Cmin(i,pos(i)),Cmin(i+1,j)+extra\_cost\}$. Then, if $p(j)<i$ we will update the variable $extra\_cost$: $extra\_cost=extra\_cost+(i-p(j))$ (this is because there are $i-p(j)$ elements which will be moved before element $i$, and their contribution would not be considered otherwise). The minimum total cost is $min\{Cmin(1,j)|1\leq j\leq n+1\}$.

The problem presented in Section 6 is a more general version of the minimum all ones problem \cite{min_all_ones}. Our dynamic programming state definition is similar to the one from \cite{min_all_ones}, but our algorithm is substantially different.

\section{Conclusions}

The novel contributions of this paper can be classified into three categories:\begin{itemize}
\item offline algorithms for (multi)point-to-(multi)point data transfer optimization
\item offline algorithms for resource processing optimization
\item the architecture of an agent-based peer-to-peer content delivery framework
\end{itemize}

The considered problems are either new, or extended (e.g. more general) versions of other existing problems. All the presented solutions make use of the specific properties of the considered problems, thus obtaining optimal or near optimal results.

The agent-based content delivery framework is focused on efficient multicast data distribution. Some of the presented offline algorithms could be used for computing a multicast data delivery tree, if information regarding the entire system is available. The framework is based on the peer-to-peer architectural model presented in \cite{p2p_mpath}, but the multicast extensions and the proposal for handling unconnectable peers are novel contributions of this paper.

As future work, we intend to consider online versions of some of the considered problems and adapt some of the offline solutions that we developed in order to obtain online techniques. These techniques could then be implemented by specialized agents for solving the problems in real-time.

\end{document}